\documentclass[twocolumn,showpacs,showkeys,preprintnumbers,amsmath,10pt]{revtex4}
\usepackage{graphicx}
\usepackage{dcolumn}
\usepackage{bm}
\begin{document}
\title{Teleportation for atomic entangled state by entanglement swapping with separate measurements in cavity QED}
\author{Zhen-Yuan Xue}
\email{zyxue@ahu.edu.cn}
\author{Ming Yang}
\author{You-Min Yi}
\author{Zhuo-Liang Cao}
\email{zlcao@ahu.edu.cn(Corresponding~author)}

\affiliation{Anhui Key Laboratory of Information Material and
Devices, School of Physics and Material Science, Anhui University,
Hefei, 230039, P R China}

\begin{abstract}
Experimentally feasible scheme for teleportation of atomic entangled
state via entanglement swapping is proposed in cavity quantum
electrodynamics (QED) without joint Bell-state measurement (BSM). In
the teleportation processes the interaction between atoms and a
single-mode nonresonant cavity with the assistance of a strong
classical driving field substitute the joint measurements. The
discussion of the scheme indicates that it can be realized by
current technologies.
\end{abstract}
 \maketitle
 \section{Introduction}
Quantum entanglement, lies at the heart of quantum information
theory (QIT), is considered to be a fundamental resource of quantum
teleportation. For a long time, it was seen merely as one of the
counterintuitive feature of quantum mechanics, import only in the
realm of Einstein-Poldolsky-Rosen (EPR) paradox \cite{1}. Only
recently has the field of quantum information begin to exploit the
applications of its novel feathers. While bipartite entangled state
is well known, multipartite entanglement is still under extensive
exploration. People soon realized that it isn't just an extension of
bipartite entanglement. For tripartite entangled quantum system, it
falls into two classes, namely, GHZ state \cite{2} and W state
\cite{3}, respectively. Now great efforts are engaged in
investigation of multipartite entanglement with its promising
features, such as, decoherence free quantum information processing,
multiparty quantum communications and so on.

Quantum teleportation, proposed by Bennett \emph{et al}. \cite{4}
and experimentally realized by Bouwmeester \emph{et al}. \cite{5}
and Boschi \emph{et al}. \cite{6}, is a process to transmit unknown
state to a remote location via a quantum channel aided by some
classical communication. It attracts extensive public attention due
to its promising applications in QIT. Recently, teleportation of
entangled states have been suggested different theoretical schemes
\cite{7,8,9,10,11,12}. But, now people pay special interest to the
physical realization of teleportation and the key steps of which are
generation of quantum channel and realization of the joint BSM.
Generation of entangled state has been realized in experiment
\cite{13,14}, but the joint BSM is very complex for experimental
realization \cite{15}, and thus been a troublemaker. In order to
avoid the difficulty, Zheng \cite{16} proposed a more feasible
teleportation scheme using resonant optical cavity without BSM and
with successful teleportation probability 0.25. But the scheme
submits to both the affections of cavity decay and the thermal
field. Then, Ye \emph{et al}. \cite{17} proposed another scheme,
also without BSM, to teleport atomic states, where the effect of
cavity decay has been eliminated and the successful probability of
teleportation is 0.5. Yang \emph{et al}. \cite{18} finally proposed
another scheme from which the affections of thermal field and cavity
decay are both eliminated.

In this paper, we consider another feasible teleportation scheme for
atomic entangled state via entanglement swapping in cavity QED.
Through analysis, we concluded that it could be succeed without BSM.
The distinct advantage of the scheme is that during the interaction
among the two atoms and the cavity field, a classical field is
simultaneously accompanied, thus the evolution of the quantum state
is independent from the state of the cavity. Thus the scheme is
insensitive to both the cavity decay and the cavity thermal field.
In addition, one BSM can be exactly converted into two separate
atomic measurements on the two relevant atoms only by one step using
the interaction between the atoms and the cavity. The rest of this
paper is organized as following: since the nonresonant optical
cavity is the main device adopted in our scheme, so in section 2, we
first calculate the evolution of the atom-cavity-field interaction
and then the generation of EPR state; section 3 is the part for
teleportation of EPR state via maximally and nonmaximally entangled
state; section 4 is the part for teleportation of \emph{N}-partite
atomic entangled GHZ class state; section 5 is the conclusion part
and remarks on the experimental feasibility of our scheme.
\section{Preparation of the quantum channel}
We consider two identical two-level atoms simultaneously interacting
with a single-mode cavity and driven by a classical field. While the
case of  $\omega_{0}=\omega$ and  $\Omega \gg\delta\gg\varepsilon$
($\Omega$ is the Rabi frequency of the classical field, $\delta$ is
the detuning between atomic transition frequency $\omega_{0}$  and
cavity frequency $\omega$, $\varepsilon$ is the coupling constant
between cavity and particle), there is no energy exchange between
the atomic system and the cavity, thus the scheme is insensitive to
both the cavity decay and the thermal field. Then the interaction
between the single-mode cavity and the atoms can be described, in
the rotating-wave approximation, as \cite{19}
\begin{eqnarray}
\label{abc}
H_{e}&=&\frac{\lambda}{2}[\sum_{j=1}^{2}(|e\rangle_{j,j}\langle e|+|g\rangle_{j,j}\langle g|)\nonumber\\
&+&\sum_{i\neq
j;j=1}^{2}(S_{j}^{+}S_{k}^{+}+S_{j}^{+}S_{k}^{-}+H.c.)]
\end{eqnarray}
where $\lambda=\varepsilon^{2}/2\delta$ and $\omega_{a}$ is the
cavity frequency; $$S_{Z}=\frac{1}{2}\sum_{j=1}^{2}
(|e\rangle_{j,j}\langle e|-|g\rangle_{j,j}\langle g|),$$
$$S_{j}^{+}=|g\rangle_{j,j}\langle e|,
S_{j}^{-}=|e\rangle_{j,j}\langle g|$$ with $|e\rangle_{j}$,
$|g\rangle_{j}$ are the excited and ground states of  $j$th atom,
respectively.

If two atoms are simultaneously sent into the cavity then they
interact with the cavity. Using the evolution operator of the system
$U(t)=\exp(-iH_{0}t)\exp(-iH_{e}t)$ with $H_{0}=\sum_{j=1}^{2}\Omega
(S_{j}^{+}+S_{j}^{-})$, $H_{e}$ is the effective Hamiltonian. it is
easy to verify the evolutions by adjusting the interaction time
$\lambda t=\pi/4$ and modulating the driving field $\Omega t=\pi$.
\begin{subequations}
\label{2}
\begin{equation}
 |g\rangle_{1}|g\rangle_{2}\rightarrow \frac{1}{\sqrt{2}}(|g\rangle_{1}|g\rangle_{2}-i|e\rangle_{1}|e\rangle_{2}),
\end{equation}
\begin{equation}
|g\rangle_{1}|e\rangle_{2}\rightarrow
\frac{1}{\sqrt{2}}(|g\rangle_{1}|e\rangle_{2}-i|e\rangle_{1}|g\rangle_{2}),
\end{equation}
\begin{equation}
|e\rangle_{1}|g\rangle_{2}\rightarrow
\frac{1}{\sqrt{2}}(|e\rangle_{1}|g\rangle_{2}-i|g\rangle_{1}|e\rangle_{2}),
\end{equation}
\begin{equation}
|e\rangle_{1}|e\rangle_{2}\rightarrow
\frac{1}{\sqrt{2}}(|e\rangle_{1}|e\rangle_{2}-i|g\rangle_{1}|g\rangle_{2}).
\end{equation}
\end{subequations}

That is to say we have prepared the four maximally bipartite
entangled states (EPR) from the reverent product state of the two
atoms. From Eq. (\ref{2}) we can see that the evolutions doesn't
depend on the cavity state, thus this scheme is insensitive to both
the cavity decay and the thermal field. Latter we will use the
states generated here to serve as quantum channels.

\section{Teleportation of bipartite entanglement}
In this section we will propose two schemes to teleport the
bipartite entangled state via entanglement swapping. The first
scheme adopts EPR state as quantum channels while the latter one
adopts two nonmaximally bipartite entangled states.

A bipartite nonmaximally entangled atomic quantum state to be
teleport can be expressed as
\begin{equation}
\label{3}
|\psi\rangle_{1}=(a|e\rangle_{1}|e\rangle_{2}+b|g\rangle_{1}|g\rangle_{2}).
\end{equation}
where $a$  and $b$ are unknown coefficients and $|a|^{2}+|b|^{2}=1$.

\subsection{Teleportation with EPR state via entanglement swapping}
Here we will teleport the bipartite entangled state by two EPR
states. Assume the two quantum channels are
\begin{subequations}
\label{eq:whole}
\begin{equation}
|\psi\rangle_{3,4}=\frac{1}{\sqrt{2}}(|ge\rangle_{3,4}-i|eg\rangle_{3,4}),
\end{equation}
\begin{equation}
|\psi\rangle_{5,6}=\frac{1}{\sqrt{2}}(|ge\rangle_{5,6}-i|eg\rangle_{5,6}),
\end{equation}
\end{subequations}
where atoms 3 and 5 belong to Alice and the rest two were sent to
Bob. Initially the quantum state of the whole system, consisting 6
atoms, is
\begin{eqnarray}
|\psi\rangle&=&\frac{1}{2}(a|e\rangle_{1}|e\rangle_{2}+b|g\rangle_{1}|g\rangle_{2})\nonumber\\
&\otimes&(|ge\rangle_{3,4}-i|eg\rangle_{3,4})\otimes(|ge\rangle_{5,6}-i|eg\rangle_{5,6}),
\end{eqnarray}
which can be rewrite as
\begin{eqnarray}
|\psi\rangle=\frac{1}{2}[|\Phi^{\pm}\rangle_{1,3}|\Phi^{\pm}\rangle_{2,5}(-b|ee\rangle\pm_{1}\pm_{2}
a|gg\rangle)_{4,6}\nonumber\\
-i|\Phi^{\pm}\rangle_{1,3}|\Psi^{\pm}\rangle_{2,5}(b|eg\rangle\pm_{1}\mp_{2}a|ge\rangle)_{4,6}\nonumber\\
-i|\Psi^{\pm}\rangle_{1,3}|\Phi^{\pm}\rangle_{2,5}(b|ge\rangle\mp_{1}\pm_{2}a|eg\rangle)_{4,6}\nonumber\\
+|\Psi^{\pm}\rangle_{1,3}|\Psi^{\pm}\rangle_{2,5}(b|gg\rangle\mp_{1}\mp_{2}a|ee\rangle)_{4,6}],
\end{eqnarray}
where$|\Phi^{\pm}\rangle=\frac{1}{\sqrt{2}}(|ee\rangle\pm
i|gg\rangle)$ and
$|\Psi^{\pm}\rangle=\frac{1}{\sqrt{2}}(|ge\rangle\pm i|eg\rangle)$
are four Bell states of atom pairs $(1,3)$ and $(2,5)$.

From the above equation we can see that, if Alice can easy
realization joint BSM on atom pairs (1,3) and (2,5), then after been
informed the measurement results, Bob can reconstruct the original
entangled state by corresponding unitary transformation, thus
achieve the goal of teleportation. But the BSM is complex for
experimental realization \cite{15}, so here we investigate separate
atomic measurements to achieve the goal of joint BSM by using
optical cavity.

We can assume, without loss of generality, the quantum state of atom
pair (1,3) is
\begin{equation}
|\Phi^{+}\rangle=\frac{1}{\sqrt{2}}(|ee\rangle+i|gg\rangle)_{1,3}
\end{equation}
Sending the atom pair to the cavity, they will simultaneously
interact with the cavity mode, by adjusting the interaction time
$\lambda t=\pi/4$ and modulating the driving field $\Omega t=\pi$,
the state of the atom pair will evolve, according to Eq. (\ref{2}),
as
\begin{eqnarray}
|\Phi^{+}\rangle=\frac{1}{\sqrt{2}}(|ee\rangle+i|gg\rangle)_{1,3}
\longrightarrow\frac{1}{2}[(|ee\rangle-i|gg\rangle)_{1,3}\nonumber\\
+i(|gg\rangle-i|ee\rangle)_{1,3}]=|e\rangle_{1}|e\rangle_{3}.
\end{eqnarray}
Similarly the other three Bell-states of the atom pairs would evolve
into
\begin{subequations}
\label{eq:whole}
\begin{equation}
|\Phi^{-}\rangle=\frac{1}{\sqrt{2}}(|ee\rangle+i|gg\rangle)_{1,3}\longrightarrow-i|g\rangle_{1}|g\rangle_{3},
\end{equation}
\begin{equation}
|\Psi^{+}\rangle=\frac{1}{\sqrt{2}}(|ge\rangle+i|eg\rangle)_{1,3}\longrightarrow|g\rangle_{1}|e\rangle_{3},
\end{equation}
\begin{equation}
|\Psi^{-}\rangle=\frac{1}{\sqrt{2}}(|ge\rangle-i|eg\rangle)_{1,3}\longrightarrow-i|e\rangle_{1}|g\rangle_{3}.
\end{equation}
\end{subequations}
In this way, we can achieve the goal of BSM with unit probability of
success by separate atomic measurements, and the four Bell-states of
atom pair $(2,5)$ can also be distinguished successfully in the same
way.

 After four separate measurements on atoms 1, 2, 3 and 5, the
entanglement of the entangled atom pairs $(1,2)$, $(3,4)$ and
$(5,6)$ disappears and the new entanglement between atoms 4 and 6 is
set up which means the entanglement swapping happens. Then after
Alice inform her measurement results to Bob; he can reconstruct the
original atomic entangled state by corresponding unitary
transformation.
\subsection{Teleportation with nonmaximally bipartite entangled state via entanglement swapping}
Here we will teleport the bipartite entangled state (\ref{3}) by two
bipartite nonmaximally entangled atomic states. Assume the quantum
channels are
\begin{subequations}
\label{eq:whole}
\begin{equation}
|\psi\rangle_{3,4}=(\alpha_{1}|ge\rangle_{3,4}-i\beta_{1}|eg\rangle_{3,4})
\end{equation}
\begin{equation}
|\psi\rangle_{5,6}=(\alpha_{2}|ge\rangle_{5,6}-i\beta_{2}|eg\rangle_{5,6})
\end{equation}
\end{subequations}
where $\alpha_{i}$  and $\beta_{i}$ are unknown coefficients and
$|\alpha_{i}|^{2}+|\beta_{i}|^{2}=1$ with $i=1,2$. We can assume
$|\alpha_{i}|>|\beta_{i}|$ without loss of generality. So initially
the quantum state of the system is
\begin{eqnarray}
|\psi\rangle_{3,4}=(a|e\rangle_{1}|e\rangle_{2}+b|g\rangle_{1}|g\rangle_{2})
(\alpha_{1}|ge\rangle_{3,4}\nonumber\\
-i\beta_{1}|eg\rangle_{3,4})
(\alpha_{2}|ge\rangle_{5,6}-i\beta_{2}|eg\rangle_{5,6})
\end{eqnarray}
which can be rewrite as
\begin{eqnarray}
|\psi\rangle&=&\frac{1}{2}[-|\Phi^{\pm}\rangle_{1,3}|\Phi^{\pm}\rangle_{2,5}(\alpha_{1}\alpha_{2}b|ee\rangle\pm_{1}\pm_{2}\beta_{1}\beta_{2}
a|gg\rangle)_{4,6}\nonumber\\
&-&i|\Phi^{\pm}\rangle_{1,3}|\Psi^{\pm}\rangle_{2,5}(\alpha_{1}\beta_{2}b|eg\rangle\pm_{1}\mp_{2}\beta_{1}\alpha_{2} a|ge\rangle)_{4,6}\nonumber\\
&-&i|\Psi^{\pm}\rangle_{1,3}|\Phi^{\pm}\rangle_{2,5}(\beta_{1}\alpha_{2}b|ge\rangle\mp_{1}\pm_{2}\alpha_{1}\beta_{2}a|eg\rangle)_{4,6}\nonumber\\
&+&|\Psi^{\pm}\rangle_{1,3}|\Psi^{\pm}\rangle_{2,5}(\beta_{1}\beta_{2}b|gg\rangle\mp_{1}\mp_{2}\alpha_{1}\alpha_{2}a|ee\rangle)_{4,6}],\nonumber\\
\end{eqnarray}
 where the notes $\pm_{i}$ and $\mp_{i}$ correspond to
the \emph{i}th BSM. We can distinguish the four Bell states by
sending the two atoms into the cavity and choosing an appropriate
interaction time. After operate separate measurements on atoms 1, 2,
3 and 5,the state of the rest two atoms will project into one of the
following state
\begin{subequations}
\label{eq:whole}
\begin{equation}
\frac{1}{2}(\alpha_{1}\alpha_{2}b|e\rangle_{4}|e\rangle_{6}\pm_{1}\pm_{2}\beta_{1}\beta_{2}a|g\rangle_{4}|g\rangle_{6}),
\end{equation}
\begin{equation}
\frac{1}{2}(\alpha_{1}\beta_{2}b|e\rangle_{4}|g\rangle_{6}\pm_{1}\mp_{2}\beta_{1}\alpha_{2}a|g\rangle_{4}|e\rangle_{6}),
\end{equation}
\begin{equation}
\frac{1}{2}(\beta_{1}\alpha_{2}b|g\rangle_{4}|e\rangle_{6}\mp_{1}\pm_{2}\alpha_{1}\beta_{2}a|e\rangle_{4}|g\rangle_{6}),
\end{equation}
\begin{equation}
\frac{1}{2}(\beta_{1}\beta_{2}b|g\rangle_{4}|g\rangle_{6}\mp_{1}\mp_{2}\alpha_{1}\alpha_{2}a|e\rangle_{4}|e\rangle_{6}).
\end{equation}
\end{subequations}
To realize the teleportation based on cavity QED, Bob must prepare
another single-mode high quality factor resonant optical cavity,
which is initially in the vacuum states  and a photon detector is
also necessary.

Bob sends one of his two atoms to the cavity and it will interact
with the cavity mode. According to the Jaynes-Cummings model, the
Hamiltonian of the resonant interaction system is
\begin{equation}
H=\omega_{a}(a^{+}a+S_{Z})+\varepsilon(aS_{+}+a^{+}S_{-}),
\end{equation}
where $a^{+}$  and  $a$ are the creation and annihilation operators
for the cavity mode, respectively. So, the initial state of the two
particles and the cavity is
\begin{equation}
|\Psi(0)\rangle_{4,6,C}=\frac{1}{2}(\alpha_{1}\alpha_{2}b|e\rangle_{4}|e\rangle_{6}-\beta_{1}\beta_{2}a|g\rangle_{4}|g\rangle_{6})|0\rangle_{C}.
\end{equation}
Without loss of generality, let Bob sends particle 6 into the
cavity, the state of the two-particle and the cavity will evolve as
\begin{eqnarray}
|\Psi(t)\rangle_{4,6,C}&=&\frac{1}{2}[(\alpha_{1}\alpha_{2}b|e\rangle_{4}(\cos\varepsilon
t|e\rangle_{6}|0\rangle_{C}-i\sin\varepsilon
t|g\rangle_{6}\nonumber\\
&\otimes&|1\rangle_{C})-\beta_{1}\beta_{2}a|g\rangle_{4}|g\rangle_{6})|0\rangle_{C}].
\end{eqnarray}
Taking the interacting time $\cos\varepsilon
t=|\beta_{1}\beta_{2}|/|\alpha_{1}\alpha_{2}|$ , we can get the
state of the quantum system after interaction as
\begin{eqnarray}
|\Psi(t)\rangle_{4,6,C}&=&\frac{1}{2}[|\beta_{1}\beta_{2}|(b|e\rangle_{4}|e\rangle_{6}
+a|g\rangle_{4}|g\rangle_{6})\nonumber\\
&\otimes&|0\rangle_{C} -i\sin\varepsilon
t|e\rangle_{4}|g\rangle_{6}|1\rangle_{C})].
\end{eqnarray}
Here we have discard the phase factor, which can be removed by a
simple rotation operator. From Eq. (17) we can see if Bob can detect
a photon in the cavity then we know our teleportation process fail.
If he can't detect any photon in the cavity, that is, the cavity is
still in the vacuum state $|0\rangle_{C}$ and the atoms are in the
state $(b|e\rangle_{4}|e\rangle_{6}+a|g\rangle_{4}|g\rangle_{6})$,
which related to the original entangled state of atoms 1 and 2 up to
a corresponding unitary transformation and by the classical
information Bob can discriminate the collapsed state and reconstruct
the original state. In other words, separate local measurements can
also achieve the goal of joint measurement and thus teleportation
with separate local measurements could succeed. We also note that
after teleportation the entanglement of atoms (1,2), (3,4) and (5,6)
disappear and the new entanglement between the atoms 4 and 6 is set
up which means the entanglement swapping do happens.

We can also calculate the probability of successful teleportation.
The other three states in Eq. (13a) and the states of (13d) have the
same successful teleportation probability, that is
$P_{a}=P_{d}=|\beta_{1}\beta_{2}|^{2}$ . As to the states (13b) and
(13c) they also have the same probability of successful
teleportation, but we have to break them into two different classes
in calculating the probability of successful teleportation. While
$|\alpha_{1}\beta_{2}|\geq |\beta_{1}\alpha_{2}|$ , the successful
teleportation probability of them is
$P_{b}=P_{c}=|\beta_{1}\alpha_{2}|^{2}$  , otherwise
$P_{b}=P_{c}=|\alpha_{1}\beta_{2}|^{2}$ . So, the total profanity of
successful teleportation in this scheme is
$P=P_{a}+P_{b}+P_{c}+P_{d}=2|\beta_{1}|^{2}$ in the case of
$|\alpha_{1}\beta_{2}|\geq |\beta_{1}\alpha_{2}|$ , otherwise
$P=2|\beta_{2}|^{2}$. We also note that if
$\alpha_{1}=\alpha_{2}=\alpha$ and $\beta_{1}=\beta_{2}=\beta$ ,
then$P_{a}=P_{d}=|\beta|^{4}$ and $P_{b}=P_{c}=|\beta\alpha|^{2}$,
thus $P=2|\beta|^{2}$ which proves to be the maximally probability
for entanglement teleportation \cite{20}.
\section{Teleportation of multipartite entanglement}
We also note that the scheme in section 3 can directly be
generalized to teleport multipartite entanglement. In this section,
we will consider the generalization of our schemes.

We first consider the former scheme, in order to teleport a
\textit{N}-atom entangled state we need \textit{N} EPR pairs, thus
the initial state of the system is
\begin{eqnarray}
|\Psi\rangle&=&(a|e\rangle_{1}\cdot\cdot\cdot|e\rangle_{n}+b|g\rangle_{1}\cdot\cdot\cdot|g\rangle_{n})\nonumber\\
&\otimes&\prod_{j=1}^{n}\frac{1}{\sqrt{2}}
(|e\rangle_{n+j}|e\rangle_{2n+j}+|g\rangle_{n+j}|g\rangle_{2n+j})
\end{eqnarray}
where atoms $1,2\cdot\cdot\cdot 2n$ belong to Alice and atoms
$(2n+1),(2n+2)\cdot\cdot\cdot (3n)$ are sent to Bob. We can
distinguish the four Bell states of atom pairs $(j,n+j)$ one by one
by using the cavity, thus can achieve unity probability and fidelity
teleport the \textit{N}-partite entangled state of atoms
$1,2\cdot\cdot\cdot n$ to another\textit{ N} atoms of
$(2n+1),(2n+2)\cdot\cdot\cdot (3n)$. After teleportation, the new
entanglement of atoms $(2n+1),(2n+2)\cdot\cdot\cdot (3n)$  is set
up.

The generalization of the latter scheme is familiar with the former
one. In order to teleport a \textit{N}-atom entangled state one need
to set up\textit{ N} bipartite nonmaximally entangled states. After
the distinguish of the four Bell-states of atom pairs $(j,n+j)$  one
by one by using the cavity, Bob only need send one of his atoms to
the resonant cavity and take a proper interaction time, if he can't
detect photon in the cavity then he can reconstruct the original
state by corresponding unitary transformation which he can know by
the aid of classical information he received from Alice.

\section{Remarks and conclusions}

Here, we will give a brief discussion on the experimental
realization of our scheme. For the large-detuned cavity, it is free
of the effects of the cavity decay and thermal field. Meanwhile it
is noted that the atomic state evolution is independent of the
cavity field state, thus based on cavity QED techniques presently
\cite{13,14,15} it might be realizable. In our scheme, the two atoms
must be simultaneously interaction with the cavity. But in real
case, we can't achieve simultaneousness in perfect precise.
Calculation on the error suggests that it only slightly affects the
fidelity of the reconstruct state \cite{21}. For the resonant
cavity, In order to realize the teleportation successfully, the
relationship between the teleportation time and the excited atom
lifetime should take into consideration. The time required to
complete the teleportation should be much shorter than that of atom
radiation. Hence, atom with a sufficiently long excited lifetime
should be chosen. For the Rydberg atoms with principal quantum
numbers 50 and 51, the interaction time is on the order is much
shorter than the atomic radiative time \cite{13}. So our scheme is
realizable by using available cavity QED techniques.

In conclusion, simple and physical schemes for teleportation of
bipartite entangled atomic state based on cavity QED are proposed.
The first scheme adopts two EPR state as quantum channels and the
latter one adopts two bipartite nonmaximally entangled states as
quantum channel. We also consider the generation of EPR state in
section 2, in the forth section we generalize the two schemes to the
case of multipartite entanglement teleportation. The presented
schemes are achieve with separate atomic measurements instead of any
types of joint measurement, which are difficult for experimental
realization. The probability of successful teleportation is obtained
and proves to be unit or maximal within current known knowledge. In
addition, our scheme is insensitive to both the thermal field and
the cavity decay, thus feasible within current cavity QED
techniques.

\begin{acknowledgments}
This work is supported by Anhui Provincial Natural Science
Foundation under Grant No: 03042401, the Key Program of the
Education Department of Anhui Province under Grant No: 2002kj029zd
and 2004kj005zd and the Talent Foundation of Anhui University.
\end{acknowledgments}

 \end{document}